# Beyond the Kolmogorov Johnson Mehl Avrami kinetics: inclusion of the spatial correlation

Authors: M. Fanfoni (1), M. Tomellini (2)
((1) Dipartimento di Fisica Università di Roma Tor Vergata, (2) Dipartimento di Scienze e Tecnologie Chimiche, Università di Roma Tor Vergata Rome Italy)

The Kolmogorov-Johnson-Mehl-Avrami model, which is a nucleation and growth poissonian process in space, has been implemented by taking into account spatial correlation among nuclei. This is achieved through a detailed study of a system of distinguishable and correlated dots (nuclei). The probability that no dots be in a region of the space has been evaluated in terms of correlation functions. The theory has been applied to describe nucleation and growth in two dimensions under constant nucleation rate, where correlation among nuclei depends upon the size of the nucleus. We propose a simple formula for describing the phase transition kinetics in the presence of correlation among nuclei. The theory is applied to the constant nucleation rate process when correlation depends upon the nucleus-birth time. It is shown that the random sequential adsorption and Tobin process can be analyzed in the framework of the simultaneous nucleation case, admitting a common rationale that is apart from an appropriate re-scaling they represent the same process from the mathematical point of view.

**1-Introduction**

It is common knowledge that in order a process in space be poissonian, dots must be dispersed at random throughout the whole space. Kolmogorov-Johnson-Mehl-Avrami [1-3] applied this process for describing phase transformations based on nucleation and growth of the new phase in the parent phase. In their model nuclei are preexisting and the nucleation rate law is given a priori and concerns all the nuclei. This means that nuclei already covered by the new phase can start growing; these nuclei were christened phantoms by Avrami. To the end of reaching a simple formula for the kinetics it is necessary to include in the computation the contribution of phantoms [4], although they apparently do not contribute to the new phase, and more than that if they did the model simply would fail. As a matter of fact there exist a class of functions that cannot be used for approximating the growth law; for instance one of these, usually employed in diffusional growths [5], is $r = t^{1/2}$, r being the linear dimension of nucleus. This limit of the KJMA model can be overcome by simply considering the actual nucleation, that is riding of the phantoms nuclei. Nevertheless this alternative point of view require a more complex mathematical treatment; since the actual nucleation is limited to the untransformed space, correlation among nuclei must be taken into account [6]. Besides it is also possible that around a growing nucleus there be a zone where nucleation is strongly reduced, for example because of the stress induced by the new phase or because of a diffusional field. In a case like this the introduction of the correlation function would be compulsory, and as witnesses by the number of papers issued since 2000 [7-12], correlated nucleation in KJMA-type phase transitions has becoming an argument of great moment.

The aim of the present paper is to give a comprehensive and a quite general formulation of some results already published [6,8,12]. Specifically the novelty are: i) the detailed demonstration of the exact kinetics of nucleation and growth in the case of non random and non simultaneous nucleation and, ii) using the result i), at the second order in the correlation functions, to evaluate the kinetics in case of correlation depending upon the nucleus size. The latter is the starting point for treating the kinetics by using the actual nucleation rate that is, in turn, accessible by experiment.

The paper is organized as follows: In section II, we describe the theory for treating a set of distinguishable classes of dots. In particular, the central point of the stochastic approach is to evaluation the probability, $Q_0$, that no dots occurs into a region of a given volume. This is achieved by exploiting the relation [13] $Q_0 = \lim_{e^{ik} \to 0} \left\langle e^{ikN} \right\rangle$, where $N$ is the stochastic variable number of dots and k is a parameter. The brackets stand for average over the distribution that will be introduced shortly. Furthermore, one can demonstrate that $\left\langle e^{ikN} \right\rangle = L[\{\exp(ik\boldsymbol{c}) - 1\}]$, in which L is an appropriate functional and $\boldsymbol{c}$ is the characteristic function of the dot classes. The bulk of calculation is confined in the Appendices. Application of the theory to KJMA-type phase transitions is presented in sect.III for correlation functions depending on nucleus size. Section IV is devoted to the simplest application of the model, that is the Dirac delta nucleation, and to the discussion of the RSA (Random Sequential Adsorption) and Tobin processes.

## II-Stochastic theory of distinguishable dots

### A- Definitions

Let us consider a countable set of distinguishable classes of dots. It goes without saying that within each class dots are indistinguishable. Moreover dots are correlated one another independently of the class they belong.

Following Van Kampen [13], each state of the sample space consists of:

a)

i) a non negative integer $m$ ;

ii) a non negative integer $s$;

iii) a m-tupla of strictly positive integers $(n_1, n_2, n_3,..., n_m)$, in such a way that their sum is equal to $s$. The m-tupla is a partition of the integer $s$.

For each $m$, $s$ and $(n_1, n_2, n_3,..., n_m)$ $s$ d-dimensional real variables exist each of them ranging the whole space:

$$\underbrace{\{\mathbf{x}_1,...,\mathbf{x}_{n_1}\}_1,\{\mathbf{x}_1,...,\mathbf{x}_{n_2}\}_2,...,\{\mathbf{x}_1,...,\mathbf{x}_{n_m}\}_m}_{s} \in \Re^{ds} \tag{1}$$

b) The probability distribution over these states is given by a sequence of nonnegative functions, $Q$, defined in the domain (1) and normalized according to

$$1 = Q_0 + \sum_{\{1\}}\sum_{s}\sum_{\Pi^s_{\{1\}}}\frac{1}{n_i!}\int Q^{(1)}_{\underset{i}{\overset{s}{p}}}d^s\mathbf{x} + \sum_{\{2\}}\sum_{s}\sum_{\Pi^s_{\{2\}}}\frac{1}{n_i!n_j!}\int Q^{(2)}_{\underset{ij}{\overset{s}{p}}}d^s\mathbf{x} + \sum_{\{3\}}\sum_{s}\sum_{\Pi^s_{\{3\}}}\frac{1}{n_i!n_j!n_k!}\int Q^{(3)}_{\underset{ijk}{\overset{s}{p}}}d^s\mathbf{x} + ... \tag{2}$$

$\{1\}$ indicates the set of all classes, $\{2\}$ the set of all distinct couples of classes, $\{3\}$ the set of all distinct terns and so on. With the adjective *distinct* we signify that, for example, the $3!$ possible terns made up by the classes $ijk$ are a unique tern. $s$ is the total number of dots, $m$ is the number of classes and then $s \geq m$ is compulsory. $\Pi^s_{\{m\}}$ is the set of all partitions of $s$ with $m$ integers. By the same token employed for classes, the set of all co-ordinates will be indicated by $\langle 1 \rangle$, the set of all distinct couples of co-ordinates by $\langle 2\rangle$ and so on. Although we are using the same symbol as average, mixing between the two is prevented by the context and by the fact that the sets of coordinates appears only as a sum index.

Let $U$ be a function on the same state space as $Q's$ of the form $U = U^{(1)} + U^{(2)} + U^{(3)} + \ldots$, where $U^{(m)} = \sum_{\{1\}}^{m} \sum_{\langle 1\rangle}^{s} u_i(\mathbf{x}_n)$ and the subscript of $u$ refers to the classes while that of x to co-ordinates, then its average over $Q's$ will be

$$\langle U\rangle = \left\langle U^{(1)}\right\rangle_{\{1\}} + \left\langle U^{(2)}\right\rangle_{\{2\}} + \left\langle U^{(3)}\right\rangle_{\{3\}} + \ldots \tag{3}$$

$$\left\langle U^{k}\right\rangle = \left\langle U^{(1)k}\right\rangle_{\{1\}} + \left\langle U^{(2)k}\right\rangle_{\{2\}} + \left\langle U^{(3)k}\right\rangle + .. \tag{4}$$

In the following we will compute $\langle U\rangle$, $\langle U^2\rangle$ and $\langle U^3\rangle$ and we will define the so-called $f$-functions and correlation functions (CFs) [13] nevertheless the bulk of calculation is reported in the appendices.

By using a short notation: $U^{(1)} = \sum_{\langle 1\rangle} u_i$, $U^{(2)} = \sum_{\langle 1\rangle} u_i + \sum_{\langle 1\rangle} u_j$. The superscript $(1)$ stands for the presence of a single class which is specified by the subscript $i$ of the single-variable function $u_i$, where the variable dependence is understood. By the same token the superscript (2) stands for the presence of two classes , namely i,j.

**B- Average quantities**

The average of $U^{(1)}$ reads

$$\left\langle U^{(1)}\right\rangle = \sum_{\{1\}} \sum_{s} \sum_{\Pi^{s}_{\{1\}}} \frac{1}{n_i!} \int Q^{(1)}_{p^s_i} \sum_{\langle 1\rangle} u_i d^s \mathbf{x} = \sum_{\{1\}} \sum_{s} \sum_{\Pi^{s}_{\{1\}}} \frac{n_i}{n_i!} \int u_i d\mathbf{x} \int Q^{(1)}_{p^s_i} d^{s-1}\mathbf{x} \tag{5}$$

which, defining the function $h_i \equiv \sum_{s} \sum_{\Pi^{s}_{\{1\}}} \frac{n_i}{n_i!} \int Q^{(1)}_{p^s_i} d^{s-1}\mathbf{x}_i$, can be rewritten as

$$\left\langle U^{(1)}\right\rangle = \sum_{\{1\}} \int u_i h_i d\mathbf{x}_i \equiv \sum_{\{1\}} (u_i h_i) \tag{6}$$

By following a similar path of computation and defining the functions $h_{i,j} \equiv \sum_s \sum_{\Pi^s_{\{2\}}} \frac{n_i}{n_i!n_j!} \int Q^{(2)}_{p^s_{ij}} d^{n_i-1}\mathbf{x}_i d^{n_j}\mathbf{x}_j$ and $h_{i,jk} \equiv \sum_s \sum_{\Pi^s_{\{3\}}} \frac{n_i}{n_i!n_j!n_k!} \int Q^{(3)}_{p^s_{ijk}} d^{n_i-1}\mathbf{x}_i d^{n_j}\mathbf{x}_j d^{n_k}\mathbf{x}_k$ one obtains

$$\left\langle U^{(2)} \right\rangle = \sum_{\{1\}} \sum_{\{1\}\setminus i} \left( u_i h_{i,j} \right) \tag{7}$$

$$\left\langle U^{(3)} \right\rangle = \sum_{\{1\}} \sum_{\{2\}\setminus i} \left( u_i h_{i,jk} \right) \tag{8}$$

where the symbol $\{m\}\setminus i$ means that all the m-tuplas are considered that do not contain the i-class and, as in eqn.5, the parenthesis denote an integration. It is now possible to define the $f$-functions for any single class, that, at odds with those defined in [13], take into account the presence of the other classes, as

$$f_i \equiv h_i + \sum_{\{1\}\setminus i} h_{i,j} + \sum_{\{2\}\setminus i} h_{i,jk} + \sum_{\{3\}\setminus i} h_{i,jk\ell} + \ldots \tag{9}$$

and thus one ends up with

$$\langle U \rangle = \sum_{\{1\}} \left( u_i f_i \right) \tag{10}$$

The meaning of the $f_i$-function is immediate: the term $f_i d\mathbf{x}_i$ gives the probability of finding a dot (any) belonging to the i class in the volume element $d\mathbf{x}_i$ around $\mathbf{x}_i$, irrespective of the location of the other dots.

As far as the evaluation of $\left\langle U^2 \right\rangle$ and $\left\langle U^3 \right\rangle$ are concerned, the computation is a more involved. However even in this case is it profitable to define $f_{ij}$ and $f_{ijk}$ functions for couples and terns of classes such that, $f_{ij} d\mathbf{x}_i d\mathbf{x}_j$ gives the probability of finding a dot of the i class in the element $d\mathbf{x}_i$ around $\mathbf{x}_i$ and a dot of the j class in the element $d\mathbf{x}_j$ around $\mathbf{x}_j$, irrespective of the location of the other dots; in a similar way $f_{ijk}$ is defined. The evaluation of the averages, which have been reported in the Appendix A, leads to the results:

$$\left\langle U^2 \right\rangle = \sum_{\{1\}} \left( u_i^2 f_i \right) + \sum_{\{1\}} \sum_{\{1\}} \left( u_i u_j f_{ij} \right) \tag{11}$$

$$\langle U^3\rangle = \sum_{\{1\}}\left(u_i^3 f_i\right) + 3\sum_{\{1\}}\sum_{\{1\}}\left(u_i^2 u_j f_{ij}\right) + \sum_{\{1\}}\sum_{\{1\}}\sum_{\{1\}}\left(u_i u_j u_k f_{ijk}\right). \tag{12}$$

**C- The functional "L"**

Let us now define the following functional of the stochastic variables $\{u\}$

$$L[\{u\}] \equiv \left\langle \prod_{\langle 1\rangle}^{n_i}(1+u_i)\prod_{\langle 1\rangle}^{n_j}(1+u_j)\prod_{\langle 1\rangle}^{n_k}(1+u_k)\ldots\right\rangle. \tag{13}$$

By making use of the averages of the previous section, eqn. 13 can be written as

$$\begin{aligned} L[\{u\}] &= 1+\sum_{\{1\}}\sum_{s}\sum_{\Pi_1^s}\frac{1}{s!}\left(u_1^s f_s\right)+\sum_{\{2\}}\sum_{s}\sum_{\Pi_2^s}\frac{1}{n_1!n_2!}\left(u_1^{n_1}u_2^{n_2} f_{n_1 n_2}\right)+\ldots = \\ &= 1+\sum_{m}\sum_{\{m\}}\sum_{s}\sum_{\Pi_m^s}\frac{1}{n_1!\ldots n_m!}\left(u_1^{n_1}\ldots u_m^{n_m} f_{n_1\ldots n_m}\right) \end{aligned} \tag{14}$$

where $f_{n_1 \ldots n_m}$ denotes the $f$-function that depends upon $s=\sum_{i=1}^{m} n_i$ variables of which $n_1$ of class 1, $n_2$ of class 2 and so on. The details of the computation are reported in Appendix C.

**D-Correlation functions**

We are now in a position to define the CFs, $g_m$, through the cluster expansion of the $f$-functions [13]. For the sake of clearness we report the case of four variables, two classes and the partition $(n_1, n_2) = (2,2)$, as follows:

$$\begin{aligned} f_{2,2}\left(1,2,\bar{3},\bar{4}\right) = {} & g_1(1)g_1(2)g_1\left(\bar{3}\right)g_1\left(\bar{4}\right)+g_2(1,2)g_2\left(\bar{3},\bar{4}\right)+g_2\left(1,\bar{3}\right)g_2\left(2,\bar{4}\right)+g_2\left(1,\bar{4}\right)g_2\left(2,\bar{3}\right)+ \\ & +g_1(1)g_3\left(2,\bar{3},\bar{4}\right)+g_1(2)g_3\left(1,\bar{3},\bar{4}\right)+g_1\left(\bar{3}\right)g_3\left(1,2,\bar{4}\right)+g_1\left(\bar{4}\right)g_3\left(1,2,\bar{3}\right)+ \\ & +g_1(1)g_1(2)g_2\left(\bar{3},\bar{4}\right)+g_1(1)g_1\left(\bar{3}\right)g_2\left(2,\bar{4}\right)+g_1(1)g_1\left(\bar{4}\right)g_2\left(2,\bar{3}\right)+ \\ & +g_1(2)g_1\left(\bar{3}\right)g_2\left(1,\bar{4}\right)+g_1(2)g_1\left(\bar{4}\right)g_2\left(1,\bar{3}\right)+g_1\left(\bar{3}\right)g_1\left(\bar{4}\right)g_2(1,2)+ \\ & +g_4\left(1,2,\bar{3},\bar{4}\right), \end{aligned} \tag{15}$$

where the bar distinguishes the class. In order to obtain the expansion in the first place we need to determine the set, $\mathsf{P}_4$, of the partition of 4, in which the generic element of the set is $(k_1, k_2, k_3, k_4)$ and the positive integers $k_i$ are determined according to the fulfillment of the relation $4 = 1k_1 + 2k_2 + 3k_3 + 4k_4$; in this case: $(4,0,0,0)$, $(0,2,0,0)$, $(1,0,1,0)$, $(2,1,0,0)$ and $(0,0,0,1)$. The first element implies four functions of one variable, the second element two functions of two variables and so on, as in eqn.(15). The second step is to form, within each element of the set $\mathsf{P}_4$, the set, $\mathrm{P}_4$, of all the permutations which give an original (independent) contribution.

The generalization to the case of $s$ variables, $m$ classes and the partition $(n_1, \ldots, n_m)$ leads to

$$f_{n_1, n_2, \ldots n_m} = \sum_{\mathsf{P}_s} \sum_{\mathrm{P}_s} \prod^{k_1} g_1 \cdots \prod^{k_n} g_n \cdots \prod^{k_s} g_s \tag{16}$$

The short notation $\prod^{k_n} g_n$ indicates the product of $k_n$ $n-$variable CFs. The $n-$variables are linked to the $m$ classes, i.e. there are $\boldsymbol{n}_1^n$ variables of class 1, $\boldsymbol{n}_2^n$ variables of class 2 and so on, in such way that

$$\sum_{j=1}^{m} \boldsymbol{n}_j^n = n \tag{17}$$

It can happen that some of the $k_n$ $m$-tuples coincides (for example in eqn.15 $g_2(1,\bar{3})$ and $g_2(2,\bar{4})$ contain the same m-tupla, i.e. the couple $\boldsymbol{n}_1^2 = 1, \boldsymbol{n}_2^2 = 1$), therefore, if $m(n/m)$ is the index that refers to the $m$-th distinct $m$-tuple satisfying eqn.(17), and $k_{m(n/m)}$ is the multiplicity of $m(n/m)$, i.e. the number of times it occurs, eqn. (16) can be written as

$$f_{n_1, n_2, \ldots n_m} = \sum_{\mathsf{P}_s} \sum_{\mathrm{P}_s} \prod_{n} \prod_{m(n/m)} [g_n(\boldsymbol{n}_{1m}^n .. \boldsymbol{n}_{mm}^n)]^{k_{m(n/m)}} \tag{18}$$

where

$$\sum_{m(n/m)}^{m} k_{m(n/m)} = k_n \tag{19}$$

and

$$\sum_{n}\sum_{m(n/m))}^{m} k_{m(n/m)} n_{jm}^{n} = n_{j} \tag{20}$$

and the notation $g_n(n_{1m}^n..n_{mm}^n)$ stands for $g_n(\mathbf{x}_1^{(1)},...,\mathbf{x}_{n_{1m}^n}^{(1)},\mathbf{x}_1^{(2)},...,\mathbf{x}_{n_{2m}^n}^{(2)}...,\mathbf{x}_1^{(m)},...,\mathbf{x}_{n_{mm}^n}^{(m)})$, where the superscript refers to the class.

The following step is to determine the number of distinct permutations. Terms are not different if they differ by the order of variables of the same class inside the individual $g_n$'s or by the order of factor $g_n$ whose $m$-tuple, $n_{1m}^n...n_{mm}^n$, pertains to the same $m(n/m)$. Thus the number of terms is

$$\frac{n_1!n_2!...n_m!}{\prod_{n}\prod_{m(n/m)}(n_{1m}^n!)^{k_{m(n/m)}}...(n_{mm}^n!)^{k_{m(n/m)}}k_{m(n/m)}!} \tag{21}$$

Inserting eqn.18 in eqn.14, and taking into account eqn.21 and the fact that $u_j^{n_j} = u_j^{(\sum_n \sum_{m(n/m)}^{m} k_{m(n/m)} n_{jm}^n)} = \prod_n \prod_{m(n/m)} u_j^{k_{m(n/m)} n_{jm}^n}$ the functional (eqn. 14) becomes

$$L[\{u\}] = 1 + \sum_{m}\sum_{\{m\}}\sum_{s}\sum_{\Pi_m^s}\sum_{P_s}\prod_{n}\prod_{m(n/m)}\frac{1}{k_{m(n/m)}!}\left[\frac{\left(u_1^{n_{1m}^n}...u_m^{n_{mm}^n} g_n(n_{1m}^n..n_{mm}^n)\right)}{n_{1m}^n!..n_{mm}^n!}\right]^{k_{m(n/m)}} \tag{22}$$

This formidable expression can be rewritten in a simpler way as

$$L[\{u\}] = \prod_{m}\prod_{\{m\}}\prod_{s}\prod_{\Pi_m^s}\sum_{k=0}^{\infty}\frac{1}{k!}\left[\frac{\left(u_1^{n_1^s}...u_m^{n_m^s} g_s(n_1^s..n_m^s)\right)}{n_1^s!..n_m^s!}\right]^{k} \tag{23}$$

As a matter of fact each term of eqn.23 can be put in one-to-one correspondence with a term of eqn.22, with the aid of constraints 17, 19, 20.

It follows that

$$L[\{u\}] = \exp\left[\sum_{m}\sum_{\{m\}}\sum_{s}\sum_{\Pi_m^s}\frac{\left(u_1^{n_1^s}...u_m^{n_m^s} g_s(n_1^s..n_m^s)\right)}{n_1^s!..n_m^s!}\right] \tag{24}$$

On the basis of eqn.24 it is possible to evaluate the probability, $Q_0$, that no dots of the i class be in the $\Delta_i$ domain, no dots of the j class be in the $\Delta_j$ domain etc. Let us denote by $\boldsymbol{c}_i$ the characteristic function of the i-class dots, defined as follows: $\boldsymbol{c}_i(\mathbf{x}) = 1$ for $\mathbf{x} \subset Ä_i$ and $\boldsymbol{c}_i(\mathbf{x}) = 0$ for $\mathbf{x} \not\subset Ä_i$. Consequently, the stochastic variable "number of dots", N, reads $N^{(m)} = \sum_{\{1\}}^{m} \sum_{\langle 1 \rangle}^{s} \boldsymbol{c}_i(\mathbf{x}_v)$. The average of the stochastic variable $e^{ikN}$, where k is a parameter, is given through eqn.13 according to $\langle \exp(ikN) \rangle \equiv L[\{\exp(ik\boldsymbol{c}) - 1\}]$. Moreover since $Q_0 = \lim_{\exp(ik)\to 0} \langle \exp(ikN) \rangle$ [13] we infer from eqn.14, the important relationship

$$Q_0 = L[\{-\boldsymbol{c}\}] \tag{25}$$

which by employing eqn.24, becomes

$$Q_0 = \exp\left[ \sum_m \sum_{\{m\}} \sum_s \sum_{\Pi_m^s} \frac{(-)^s \int_{\Delta_1} d^{n_1^s}\mathbf{x}_1 ... \int_{\Delta_m} d^{n_m^s}\mathbf{x}_m g_s \left( \boldsymbol{n}_1^s .. \boldsymbol{n}_m^s \right)}{\boldsymbol{n}_1^s ! .. \boldsymbol{n}_m^s !} \right] \tag{26}$$

**E- Continuum limit**

In order to perform the continuum limit of eqn.26 it is convenient to rewrite the CFs according to [14] as:

$$g_s(\boldsymbol{n}_1^s, \boldsymbol{n}_2^s, ..., \boldsymbol{n}_m^s) = n_1^{n_1^s} n_2^{n_2^s} ... n_m^{n_m^s} \tilde{g}_s \tag{27}$$

where $n_1,...n_m$ are the densities of dots of classes 1,..,m, respectively. Moreover it is poissible to show that eqn. 26 can be rewritten as:

$$Q_0 = \exp\left[ \sum_s \frac{(-)^s}{s!} \sum_{i_1} ... \sum_{i_s} (n_{i_1} ... n_{i_s}) \int_{\Delta_{i_1}} d\mathbf{x}^{(i_1)} ... \int_{\Delta_{i_s}} d\mathbf{x}^{(i_s)} \tilde{g}_s(\mathbf{x}^{(i_1)}, ..., \mathbf{x}^{(i_s)}) \right] \tag{28}$$

where the multiplicity $(\boldsymbol{n}_i^s)$ is naturally included by reason of the fact that the sum indexes $(i_k)$ run, independently, over all classes. In the Appendix C the equivalence of eqn.26 and 28 is shown for the case of three classes. The continuum limit implies: $n_i \to \Delta n(t_i) = I(t_i)\Delta t$, $\Delta n(t_i)$ is the number of dots between the $t_i$ and $t_i+\Delta t$ classes;thus eqn.28 becomes

$$Q_0 = \exp\left\{\sum_s \frac{(-)^s}{s!}\int_0^t dt_1 \int_0^t dt_2 ... \int_0^t dt_s I(t_1)...I(t_s) \int_{\Delta_{t_1}} d\mathbf{x}_1 ... \int_{\Delta_{t_s}} d\mathbf{x}_s \tilde{g}_s(\mathbf{x}_1...\mathbf{x}_s)\right\} \quad (29)$$

It is worth noticing that in case of nucleation into an homogeneous medium, $\tilde{g}_1 = 1$ and the contribution at s=1, in the argument of eqn.29, gives $-X_e = -\int_0^t dt' I(t')|\Delta_{t'}|$ where $|\Delta_{t'}|$ is the measure of the domain and $X_e$ is the so called extended volume in dD. The kinetics can be rewritten according to:

$$Q_0 = \exp(-\gamma X_e) \quad (30)$$

where

$$\gamma = 1 - \sum_s \frac{(-)^s}{s! X_e}\int_0^t dt_1 \int_0^t dt_2 ... \int_0^t dt_s I(t_1)...I(t_s) \int_{\Delta_{t_1}} d\mathbf{x}_1 ... \int_{\Delta_{t_s}} d\mathbf{x}_s \tilde{g}_s(\mathbf{x}_1..\mathbf{x}_s) \quad (31)$$

Correlation among nuclei is entirely considered through the γ term and, in the limit γ=1 eqn.29 reduces to the one of the poisson process.

### III- KJMA-type phase transitions case

The foregoing results can be exploited for describing phase transition kinetics in which nucleation and growth laws are given a priori (JMAK-type transitions). Although the theory is independent of the space dimension, for the sake of simplicity we will deal with the 2D case. To this end $I(t)$ becomes the nucleation rate, $|\Delta_{t'}| = \pi R^2(t-t')$, being $R(t-t')$ the growth law and $(1-Q_0)$ is equal to the fraction of transformed phase (surface).

Two cases can be pursued, they are:

a) the correlation functions do not depend on the birth time of the nuclei;

b) the correlation function depends on the birth time of the nuclei.

The simplest example of correlation is the "hard core" case. For the a-b cases, as far as the lowest order term of $g_2$ is concerned, the two dots correlation functions are, respectively,

I) $\tilde{g}_2^{(0)} = H(r - R_{hc}) - 1$ (32)

II) $\tilde{g}_2^{(0)}(i,j) = H[r - (R_{ij} + \rho)] - 1$ (33)

where H is the Heaviside function, $r = |\mathbf{r}_1 - \mathbf{r}_2|$ is the modulus of the distance between nuclei 1 and 2, $R_{hc}$ is the hard core distance, $R_{ij} = R(t_i - t_j)$ and ρ is, in general, a function of time. In other words around each island a region exists where nucleation is forbidden. The first case (I) has been thoroughly analyzed in ref.1 by only considering terms up to s=2 for constant nucleation rate and linear growth law. In particular an analytical expression of the γ exponent has been given by de-coupling the integral on the $g_2$ function in eqn.31 once expressed through polar coordinates [1]. It is

$$g(S_e, S^*) = 1 + \frac{1}{2}\left\{ S_e H\left(1 - \frac{3S_e}{S^*}\right) + \left[S^* - 2S_e (S^*/3S_e)^{3/2}\right] H\left(\frac{3S_e}{S^*} - 1\right)\right\} \tag{34}$$

where $S^* = \pi R_{hc}^2 I (3S_e / a^2 \pi I)^{1/3}$, $I$ and $a$ being the nucleation and growth rates respectively. In this case the size of the capture zone is constant and it is the same for each nucleus. Conversely, the second case is indeed more realistic since the size of the capture zone grows with the nucleus. This problem has been dealt with firstly in ref. [10,11] employing a different approach which is not based on the use of CFs. The formalism developed in section II is general and can be applied to case II as well. By retaining terms up to s=2 and considering a constant nucleation rate, eqn. 31 reads

$$g = 1 + \frac{S_e}{2} - \frac{I^2}{2S_e} \int_0^t dt' \int_0^t dt'' \Gamma[R(t,t'), R(t,t'')] \tag{35}$$

where, in case of linear growth,

$$\Gamma[R', R''] = 2\pi \int_0^{R''} r\,dr \int_0^{2\pi} d\theta \int_0^{\eta(R',r,\theta)} \psi H[\psi - (R' - R'' + \rho)] d\psi \tag{36}$$

where $\eta(R', r, \theta) = -r\cos\theta + (R'^2 - r^2 \sin^2\theta)^{1/2}$, ρ is considered as a constant and $R'' < R'$.

The calculation points out that the contribution of eqn.36 is numerically unimportant for $r > 0.7\left(\frac{a}{I}S_e\right)^{1/3}$ which, for Se=3, that is its typical maximum value, gives $r > \left(\frac{a}{I}\right)^{1/3}$. So the following expression for the kinetics can be retained:

$$S \cong 1 - \exp\left(-S_e - \frac{S_e^2}{2}\right) \qquad (37)$$

We conjecture that eqn.37 can be employed whenever nuclei are correlated according to eqn.33. Moreover in eqn.37 the extended surface is computed using the actual nucleation rate which is an experimentally accessible quantity.

Apparently the aforementioned nucleation rate is unphysical the number of nucleation events being constant for each value of the surface fraction available for nucleation. A more suitable choice would require $I(t) = IS_{free}$ where $S_{free}$ is the surface fraction available for nucleation.

It is worth pointing out that in case I), when the radius of the nucleus becomes larger than $R_{hc}$, the probability to nucleate a phantom is different from zero. On the contrary in case II) phantoms cannot nucleate at all. Furthermore, in the limiting case ρ=0 the S(t) kinetics given by eqns.30, 35-36 reduces to the KJMA solution, but evaluated by using the actual nucleation rate. To be specific, the phase transition, analyzed as a problem of correlated nucleation (nucleation is not permitted on the already transformed surface) with constant value of the actual nucleation rate, $I$, is stochastically equivalent to a KJMA problem for the non constant nucleation rate $I(t) = \frac{I}{(1-S)}$. In other words the difference between the two view points lays in the inclusion of phantoms as required in the KJMA treatment of phase transitions.

## IV- Simultaneous nucleation

### A-Theory

The case of simultaneous nucleation, that is when all nuclei start growing at the same time, can be also dealt with by simply considering a single class of nuclei. Under these circumstances the normalization condition of the Q's, eqn.2, reduces to

$$1 = Q_0 + \sum_s \frac{1}{s!}\int Q_s(\mathbf{x}_1,...,\mathbf{x}_s)d^s\mathbf{x} \qquad (38)$$

The computation of the $\langle U\rangle$, $\langle U^2\rangle$, etc averages permits a straightforward definition of the f-functions according to $f_n = h_n = \sum_s \frac{1}{(s-n)!}\int Q_s d^{s-n}\mathbf{x}$, where n denotes the number of dots. By the same token the functional eqn.13 and eqn.14 become, respectively $L[\{u\}] = \left\langle \prod_{\langle 1\rangle}(1+u)\right\rangle$ and $L[\{u\}] = 1 + \sum_{n=1}^{\infty}\frac{1}{n!}\left(u(\mathbf{x}_1)...u(\mathbf{x}_n)f_n\right)$. Furthermore, the probability that no dots be in a domain, say $\Delta$ , is eventually evaluated according to

$$Q_0(\Delta) = L[\{-c\}] = 1 + \sum_n \frac{(-)^n}{n!}\int_\Delta f_n d^n\mathbf{x} \tag{39}$$

where $c$ is the characteristic function of the class. By employing the cluster expansion of the f-function, one can switch from f-function to the g-CFs description. Eqn. 39 can be recast in the form

$$Q_0(\Delta) = \exp\left(\sum_{n=1}^{\infty}\frac{(-)^n}{n!}\int_\Delta g_n d^n\mathbf{x}\right) \tag{40}$$

It is worth pointing out that in eqn.39 as many as infinite terms of the cluster expansion contribute to a single exponential term in eqn.40. Furthermore, the latter can be equally obtained by substituting in the continuum limit eqn.29 the nucleation rate $I(t) = N\delta(t)$, δ being Dirac's delta function.

Eqn. 40 can be employed for computing the fraction of transformed surface, S, in phase transition ruled by simultaneous nucleation and growth processes. To this end let us consider disk shaped nuclei of radius R; overlap among disks is allowed. In fact, as far as the collision mechanism among nuclei is concerned, disk overlap mimics a phase transformation governed by the impingement process. The transformed surface is

$$S = 1 - Q_0(\Delta_R) = 1 - Q^g_{0(P)}(\Delta_R) \tag{41}$$

with

$$g = 1 - \left( \sum_{n=2}^{\infty} \frac{(-)^n}{S_e n!} \int_{\Delta_R} g_n d^n \mathbf{x} \right) \qquad (42)$$

In eqn.41 $|\Delta_R| = pR^2$, is the measure of the domain, namely the nucleus area in case of unimpeded growth, $S_e = N|\Delta_R|$, $N = f_1 = g_1$ being the nucleation density and $Q_{0(P)}(\Delta_R) = \exp(-S_e)$ the KJMA kinetics. It goes without saying that in real growth R is a function of time i.e. the nucleus growth law.

In reference 8 eqn.41 has been applied for describing 2D transitions driven by correlated nucleation according to the hard disk model. Specifically, nucleation is considered to be precluded into a circular region, of radius $R_{hc}$, centered around each nucleus. The fraction of covered space has been evaluated by retaining terms up to the second order in the correlation function, according to the hard disk model (case I sect.III). The stochastic approach, eqns.41,42, considering only the $g_2^{(0)}$ contribution, yields an excellent description of the MC output up to $S^* = pNR_{hc}^2$ =0.7-1. The accuracy of the description gets worse as the correlation increases, because the contribution of correlation functions of order higher than 2 becomes more and more important. An analytical expression of the γ exponent has been also derived by de-coupling, in eqn.41, the integral on the radial distribution function once expressed through polar coordinates. The overestimation of the integration domain results in an overestimation of the value of γ, yet, for a fortuitous case, this overestimation compensates surprisingly well for the neglected terms in the $g_m$ series, leading to a very good description of the kinetics in a rather wide range of S* values. The approximate expression for γ is

$$g = 1 + \frac{1}{2}\left[ S_e H\left(1 - \frac{S_e}{S^*}\right) + S^* H\left(\frac{S_e}{S^*} - 1\right)\right] \qquad (43)$$

**B- Random Sequential Adsorption and the Tobin processes**

This section is devoted to the discussion of two processes which, although seemingly different, admit a common rationale in the framework of the aforementioned theory: the continuous-space RSA and the Tobin process. The RSA process consists in throwing disks, at random, on a surface where overlap among disks is not allowed. Such a process models the random non-ideal adsorption of molecules at a solid surface when interaction among molecules behaves according to the hard-disk potential. Tobin process has been formulated in 1974 [15] in the ambit of thin film growth and refers to the model case of non simultaneous nucleation and instantaneous growth up to a finite value of the disk radius. Tobin's process is the equivalent of throwing disks at random onto a flat surface removing any disks whose center falls into an area occupied by previously thrown disks. The kinetic problem of both RSA and Tobin's processes consists in determining the fraction of substrate surface covered by molecules and by film, respectively.

As far as the RSA kinetics is concerned we note that eqn.40 can be employed for computing the adsorption rate of disks of diameter $\sigma$. In fact the adsorption rate is

$$\frac{dN}{dt} = FQ_0(\Delta_s) \tag{44}$$

where F is the flux of disks which impinge at the surface, $Q_0(\Delta_s)$ is the probability that an incoming disk find enough room to be adsorbed and $|\Delta_s| = \pi\sigma^2$. It is evident that this is nothing but a problem of correlated simultaneous nucleation according to a hard disk model with $R_{hc} = \sigma$. Consequently, in terms of the previously defined $S_e$ and S* quantities, the continuous-space RSA implies $S_e = S^* = N|\Delta_s| = \pi N\sigma^2$ where the surface coverage of adsorbed disk is

$$S = N\pi\sigma^2/4. \tag{45}$$

By employing eqn.43 for the $\gamma$ exponent one ends up with

$$\frac{dN}{dt} = F\exp[-4S(1+2S)] \tag{46}$$

and

$$\frac{dS}{d\tau} = \exp[-4S(1+2S)] \tag{47}$$

where $\tau = \pi\sigma^2 Ft/4$. As shown in ref.8 the adsorption rate eqn.47 is in very good agreement with the MC simulation, except in the high coverage regime that is characterized by very low adsorption rates. Eqn.47 is not suitable for studying the whole behavior of the kinetics, indeed it does not predict any maximal coverage. This issue is better tackled by employing f-function representation of the Q probability eqn.39 for in this case only terms up to n=5 has to be retained [16]. The expression for the $dS/d\tau$ rate is a polynomial and the adsorption process is therefore characterized by an asymptotic coverage value, usually called jamming point. For instance, the expansion of $dS/d\tau$ in terms of f-functions up to the second order term $f_2$ by considering only the Heaviside contribution to the radial distribution function gives

$$\frac{dS}{d\tau} \approx 1-4S+8(1-\alpha)S^2 \qquad (48)$$

where $\alpha = 1-3^{3/2}/4\pi$ is referred to as impingement factor [17,6]. However, truncation of the series to the second order term is not accurate for determining the jamming point since eqn.48 vanishes at $S = 0.35$ whereas the saturation coverage occurs at about 0.547 [18].

Let us now pass to consider Tobin's process. The constraint to which the nucleation event is subjected clearly indicates that, also in this case, we are facing a hard core correlation problem with $R_{hc} \equiv R = \sigma/2$, σ being the disk (i.e. nucleus) diameter. The fraction of transformed surface is therefore given by eqn.41 where the extended surface, $S_e$, is now

$$S_e(t) = \pi N(t)\sigma^2/4 \qquad (49)$$

$N(t)$ being the disk density at time t. It is worth noticing the analogies and differences between RSA and Tobin's processes: i) hard core radius differs by a factor of 2; ii) in Tobin's process, at variance with the RSA, due to the overlap among disks the fraction of transformed surface is not equal to the extended surface $S_e$. The application of eqn.41 for describing Tobin's process has been thoroughly discussed in ref.6 also in relation to the kinetic solutions previously proposed in the literature. It is worth noticing that for both processes the $f_n$ function with n>5 are identically nil. Consequently both the RSA rate and the fraction of covered surface in Tobin's process admit a saturation point, that is $dS/d\tau$ and S are zero at finite values of $S$ and $S_e$, respectively, which are the zeroes of the $Q_0$ polynomial. In summary Tobin's process for disks of radius $R = \sigma$ is stochastically "equivalent" to a RSA process of disks of radius $R = \sigma/2$.

## Conclusions

The exact solution for the kinetics of growth of spatially correlated nuclei for any nucleation function has been demonstrated in section II. It has been applied to model phase transition kinetics driven by a constant actual nucleation rate, when the correlation function depends on the birth time of the nucleus through the characteristic distance $r$. In this case phantoms do not exist and the kinetics reduces to eqn.37 where the extended surface is computed on the actual nucleation rate. In fact, the same kinetics has been also obtained for Tobin process discussed in section IV-B. We conjecture that, in order to describe the kinetics of hard core correlated nuclei (eqn.33), the KJMA formula must be substituted by eqn.37. Work is in progress to substantiate the conjecture.

**Appendix A**

The purpose of this appendix is to derive eqns. 11 and 12. In the first place we evaluate

$$\left\langle U^2\right\rangle = \left\langle U^{(1)2}\right\rangle_{\{1\}} + \left\langle U^{(2)2}\right\rangle_{\{2\}} + \left\langle U^{(3)2}\right\rangle_{\{3\}} + \ldots \tag{A1}$$

Let us begin expanding the three terms of series (A1)

$$U^{(1)2} = \sum_{\langle 1\rangle} u_i \sum_{\langle 1\rangle} u_i = \sum_{\langle 1\rangle} u_i^2 + 2\sum_{\langle 2\rangle} u_i u_i \; ; \tag{A2}$$

$$\begin{aligned} U^{(2)2} &= \left(\sum_{\langle 1\rangle} u_i + \sum_{\langle 1\rangle} u_j\right)^2 = \left(\sum_{\langle 1\rangle} u_i\right)^2 + (j)^2 + 2\sum_{\langle 1\rangle} u_i \sum_{\langle 1\rangle} u_j = \\ &= \left(\sum_{\langle 1\rangle} u_i^2 + 2\sum_{\langle 2\rangle} u_i u_i\right) + (j) + 2\sum_{\langle 1\rangle} u_i \sum_{\langle 1\rangle} u_j ; \end{aligned} \tag{A3}$$

$$\begin{aligned} U^{(3)2} &= \left(\sum_{\langle 1\rangle} u_i + \sum_{\langle 1\rangle} u_j + \sum_{\langle 1\rangle} u_k\right)^2 = \left(\sum_{\langle 1\rangle} u_i\right)^2 + (j) + (k) + 2\sum_{\langle 1\rangle} u_i \sum_{\langle 1\rangle} u_j + (ik) + (jk) = \\ &= \left(\sum_{\langle 1\rangle} u_i^2 + 2\sum_{\langle 2\rangle} u_i u_i\right) + (j) + (k) + 2\sum_{\langle 1\rangle} u_i \sum_{\langle 1\rangle} u_j + (ik) + (jk). \end{aligned} \tag{A4}$$

The addends made up by a letter (class index) in parenthesis denote, in a very short way, the same term on their left in which the class index is that in parenthesis. Regarding the evaluation of the averages they reduce to

$$\left\langle U^{(1)2}\right\rangle_{\{1\}} = \sum_{\{1\}} \left(u_i^2 h_i\right) + \sum_{\{1\}} \left(u_i u_i h_{ii}\right) \tag{A5}$$

where $h_{ii} \equiv \sum_s \sum_{\Pi_i^s} \frac{n_i (n_i - 1)}{n_i !} \int Q_{p_i^s}^{(1)} d^{s-2} \mathbf{x}_i$ ,

$$\left\langle U^{(2)2}\right\rangle_{\{2\}} = \sum_{\{1\}} \sum_{\{1\}\setminus i} \left(u_i^2 h_{i,j}\right) + \sum_{\{1\}} \sum_{\{1\}\setminus i} \left(u_i u_i h_{ii,j}\right) + 2\sum_{\{2\}} \left(u_i u_j h_{ij}\right) \tag{A6}$$

where $h_{ii,j} \equiv \sum_s \sum_{\Pi^s_{\{2\}}} \frac{n_i(n_i-1)}{n_i!n_j!} \int Q^{(2)}_{p^s_{ij}} d^{n_i-2}\mathbf{x}_i d^{n_j}\mathbf{x}_j$ and $h_{ij} \equiv \sum_s \sum_{\Pi^s_{\{2\}}} \frac{n_i n_j}{n_i!n_j!} \int Q^{(2)}_{p^s_{ij}} d^{n_i-1}\mathbf{x}_i d^{n_j-1}\mathbf{x}_j$.

Since $2\sum_{\{2\}}(u_i u_j h_{ij}) = \sum_{\{1\}} \sum_{\{1\}\setminus i}(u_i u_j h_{ij})$, eqn.A6 can be also written as

$$\left\langle U^{(2)2} \right\rangle_{ij} = \sum_{\{1\}} \sum_{\{1\}\setminus i} \left[ \left(u_i^2 h_{i,j}\right) + \left(u_i u_i h_{ii,j}\right) + \left(u_i u_j h_{ij}\right) \right] \qquad \text{(A7)}$$

As far as $\left\langle U^{(3)2} \right\rangle_{\{3\}}$ is concerned

$$\left\langle U^{(3)2} \right\rangle_{\{3\}} = \sum_{\{3\}} \left[ \left(u_i^2 h_{i,jk}\right) + (j) + (k) + \left(u_i u_i h_{ii,jk}\right) + (j) + (k) \right] + 2\sum_{\{3\}} \left[ \left(u_i u_j h_{ij,k}\right) + (ki) + (jk) \right] =$$
$$= \sum_{\{1\}} \sum_{\{2\}\setminus i} \left[ \left(u_i^2 h_{i,jk}\right) + \left(u_i u_i h_{ii,jk}\right) \right] + \sum_{\{1\}} \sum_{\{1\}\setminus i} \sum_{\{1\}\setminus i.j} \left(u_i u_j h_{ij,k}\right) \qquad \text{(A8)}$$

or

$$\left\langle U^{(3)2} \right\rangle_{\{3\}} = \sum_{\{1\}} \sum_{\{2\}\setminus i} \left[ \left(u_i^2 h_{i,jk}\right) + \left(u_i u_i h_{ii,jk}\right) \right] + \sum_{\{1\}} \sum_{\{1\}\setminus i} \sum_{\{1\}\setminus i.j} \left(u_i u_j h_{ij,k}\right) \qquad \text{(A9)}$$

where $h_{ii,jk} \equiv \sum_s \sum_{\Pi^s_{\{3\}}} \frac{n_i(n_i-1)}{n_i!n_j!n_k!} \int Q^{(3)}_{p^s_{ijk}} d^{n_i-2}\mathbf{x}_i d^{n_j}\mathbf{x}_j d^{n_k}\mathbf{x}_k$ and $h_{ij,k} \equiv \sum_s \sum_{\Pi^s_{\{3\}}} \frac{n_i n_j}{n_i!n_j!n_k!} \int Q^{(3)}_{p^s_{ijk}} d^{n_i-1}\mathbf{x}_i d^{n_j-1}\mathbf{x}_j d^{n_k}\mathbf{x}_k$.

Combining eqns. A5, A7, A9 the average of $U^2$ is at last achieved

$$\left\langle U^2 \right\rangle = \sum_{\{1\}}\left(u_i^2 h_i\right) + \sum_{\{1\}}\left(u_i u_i h_{ii}\right) + \sum_{\{1\}} \sum_{\{1\}\setminus i} \left[ \left(u_i^2 h_{i,j}\right) + \left(u_i u_i h_{ii,j}\right) + \left(u_i u_j h_{ij}\right) \right] +$$
$$+ \sum_{\{1\}} \sum_{\{2\}\setminus i} \left[ \left(u_i^2 h_{i,jk}\right) + \left(u_i u_i h_{ii,jk}\right) \right] + \sum_{\{1\}} \sum_{\{1\}\setminus i} \sum_{\{1\}\setminus i.j} \left(u_i u_j h_{ij,k}\right) =$$
$$= \sum_{\{1\}} \left[ \left( u_i^2 \left\{ h_i + \sum_{\{1\}\setminus i} h_{i,j} + \sum_{\{2\}\setminus i} h_{i,jk} + \ldots \right\} \right) + \left( u_i u_i \left\{ h_{ii} + \sum_{\{1\}\setminus i} h_{ii,j} + \sum_{\{2\}\setminus i} h_{ii,jk} + \ldots \right\} \right) + \sum_{\{1\}\setminus i} \left( u_i u_j \left\{ h_{ij} + \sum_{\{1\}\setminus i.j} h_{ij,k} \right\} \right) \right]$$

(A10)

By defining the $f$ -functions for any couple of classes as

$$f_{ii} \equiv h_{ii} + \sum_{\{1\}\backslash i} h_{ii,j} + \sum_{\{2\}\backslash i} h_{ii,jk} + \sum_{\{3\}\backslash i} h_{ii,jk\ell} + \ldots \tag{A11}$$

$$f_{ij} \equiv h_{ij} + \sum_{\{1\}\backslash i,j} h_{ij,k} + \ldots \tag{A12}$$

eqn.A10 finally gives

$$\left\langle U^2 \right\rangle = \sum_{\{1\}} \left( u_i^2 f_i \right) + \sum_{\{1\}} \sum_{\{1\}} \left( u_i u_j f_{ij} \right) \tag{A13}$$

Concerning the calculation of

$$\left\langle U^3 \right\rangle = \left\langle U^{(1)^3} \right\rangle_{\{1\}} + \left\langle U^{(2)^3} \right\rangle_{\{2\}} + \left\langle U^{(3)^3} \right\rangle_{\{3\}} + \ldots , \tag{A14}$$

let us begin expanding the first three terms of the series (A14):

$$U^{(1)^3} = \sum_{\langle 1 \rangle} u_i \sum_{\langle 1 \rangle} u_i \sum_{\langle 1 \rangle} u_i = \sum_{\langle 1 \rangle} u_i^3 + 3 \sum_{\langle 2 \rangle} u_i^2 u_i + 3! \sum_{\langle 3 \rangle} u_i u_i u_i ; \tag{A15}$$

$$\begin{aligned} U^{(2)^3} &= \left( \sum_{\langle 1 \rangle} u_i + \sum_{\langle 1 \rangle} u_j \right)^3 = \left( \sum_{\langle 1 \rangle} u_i \right)^3 + (j) + 3 \left( \sum_{\langle 1 \rangle} u_i \right)^2 \sum_{\langle 1 \rangle} u_j + (ji) = \\ &= \left( \sum_{\langle 1 \rangle} u_i^3 + 3 \sum_{\langle 2 \rangle} u_i^2 u_i + 3! \sum_{\langle 3 \rangle} u_i u_i u_i \right) + (j) + 3 \left( \sum_{\langle 1 \rangle} u_i^2 + 2 \sum_{\langle 2 \rangle} u_i u_i \right) \sum_{\langle 1 \rangle} u_j + (ji); \end{aligned} \tag{A16}$$

$$U^{(3)3} = \left(\sum_{\langle 1\rangle} u_i + \sum_{\langle 1\rangle} u_j + \sum_{\langle 1\rangle} u_k\right)^3 = \left(\sum_{\langle 1\rangle} u_i\right)^3 + (j) + (k) + 3\left(\sum_{\langle 1\rangle} u_i\right)^2 \sum_{\langle 1\rangle} u_j + (ik) + (jk) +$$
$$+3\sum_{\langle 1\rangle} u_i \left(\sum_{\langle 1\rangle} u_j\right)^2 + (ik) + (jk) + 6\sum_{\langle 1\rangle} u_i \sum_{\langle 1\rangle} u_j \sum_{\langle 1\rangle} u_k =$$
$$= \left(\sum_{\langle 1\rangle} u_i^3 + 3\sum_{\langle 2\rangle} u_i^2 u_i + 3!\sum_{\langle 3\rangle} u_i u_i u_i\right) + (j) + (k) +$$
$$+3\left(\sum_{\langle 1\rangle} u_i^2 + 2\sum_{\langle 2\rangle} u_i u_i\right)\sum_{\langle 1\rangle} u_j + (ik) + (jk) +$$
$$+3\sum_{\langle 1\rangle} u_i \left(\sum_{\langle 1\rangle} u_j^2 + 2\sum_{\langle 2\rangle} u_j u_j\right) + (ik) + (jk) + \tag{A17}$$
$$+6\sum_{\langle 1\rangle} u_i \sum_{\langle 1\rangle} u_j \sum_{\langle 1\rangle} u_k .$$

The averages are:

$$\left\langle U^{(1)3} \right\rangle_{\{1\}} = \sum_{\{1\}} \left(u_i^3 h_i\right) + 3\sum_{\{1\}} \left(u_i^2 u_i h_{ii}\right) + \sum_{\{1\}} \left(u_i u_i u_i h_{iii}\right) \tag{A18}$$

where $h_{iii} = \sum_s \sum_{\Pi_i^s} \frac{n_i(n_i-1)(n_i-2)}{n_i!} \int Q^{(1)}_{p_i^s} d^{s-3}\mathbf{x}_i$;

$$\left\langle U^{(2)3} \right\rangle_{\{2\}} = \sum_{\{1\}} \sum_{\{1\}\setminus i} \left[\left(u_i^3 h_{i,j}\right) + 3\left(u_i^2 u_i h_{ii,j}\right) + \left(u_i u_i u_i h_{iii,j}\right)\right] + \left\langle 3\left(\sum_{\langle 1\rangle} u_i^2 + 2\sum_{\langle 2\rangle} u_i u_i\right)\sum_{\langle 1\rangle} u_j + (ji)\right\rangle_{\{2\}} \tag{A19}$$

and the computation of the last two terms leads to

$$\left\langle 3\sum_{\langle 1\rangle} u_i^2 \sum_{\langle 1\rangle} u_j + (ji)\right\rangle_{\{2\}} = 3\sum_{\{2\}} \left[\left(u_i^2 u_j h_{ij}\right) + \left(u_j^2 u_i h_{ji}\right)\right] = 3\sum_{\{1\}} \sum_{\{1\}\setminus i} \left(u_i^2 u_j h_{ij}\right) \tag{A20}$$

and

$$\left\langle 3 \cdot 2\sum_{\langle 2\rangle} u_i u_i \sum_{\langle 1\rangle} u_j + (ji)\right\rangle_{\{2\}} = 3\sum_{\{2\}} \left[\left(u_i u_i u_j h_{iij}\right) + \left(u_j u_j u_i h_{jji}\right)\right] = 3\sum_{\{1\}} \sum_{\{1\}\setminus i} \left(u_i u_i u_j h_{iij}\right) \tag{A21}$$

Combining eqns.A19-A21 the average $\left\langle U^{(2)3}\right\rangle_{\{2\}}$ reduce to

$$\left\langle U^{(2)3}\right\rangle_{\{2\}} = \sum_{\{1\}}\sum_{\{1\}\backslash i}\left[\left(u_i^3 h_{i,j}\right)+3\left(u_i^2 u_i h_{ii,j}\right)+\left(u_i u_i u_i h_{iii,j}\right)+3\left(u_i^2 u_j h_{ij}\right)+3\left(u_i u_i u_j h_{iij}\right)\right] \qquad \text{(A22)}$$

For the sake of clarity, in calculating $\left\langle U^{(3)3}\right\rangle$ it is convenient to proceed considering the last term of eqn.A17 line by line.

The three sums over $\langle 1\rangle$, $\langle 2\rangle$ and $\langle 3\rangle$ of the first line lead, considering also the contribution of $(j)$ and $(k)$, respectively, to $\sum_{\{1\}}\sum_{\{2\}\backslash i}\left(u_i^3 h_{i,jk}\right)$, $3\sum_{\{1\}}\sum_{\{2\}\backslash i}\left(u_i^2 u_i h_{ii,jk}\right)$ and $\sum_{\{1\}}\sum_{\{2\}\backslash i}\left(u_i u_i u_i h_{iii,jk}\right)$. To obtain these relations it has been exploited the equality: $\sum_{\{3\}}\left[(i,jk)+(j,ik)+(k,ij)\right]=\sum_{\{1\}}\sum_{\{2\}\backslash i}(i,jk)$.

Therefore the contribution of the entire first line is

$$\sum_{\{1\}}\sum_{\{2\}\backslash i}\left[\left(u_i^3 h_{i,jk}\right)+3\left(u_i^2 u_i h_{ii,jk}\right)+\left(u_i u_i u_i h_{iii,jk}\right)\right] \qquad \text{(A23)}$$

Concerning the second line, let us begin considering the average of $3\sum_{\langle 1\rangle}u_i^2\sum_{\langle 1\rangle}u_j$ and of the analogous terms $(ik)$ e $(jk)$; they lead to $3\sum_{\{3\}}\left[\left(u_i^2 u_j h_{ij,k}\right)+\left(u_i^2 u_k h_{ik,j}\right)+\left(u_j^2 u_k h_{jk,i}\right)\right]$ and from the same terms in the third line one gets $3\sum_{\{3\}}\left[\left(u_j^2 u_i h_{ji,k}\right)+\left(u_k^2 u_i h_{ki,j}\right)+\left(u_k^2 u_j h_{kj,i}\right)\right]$. Since $\left(u_i^2 u_j h_{ij,k}\right)=\left(u_j^2 u_i h_{ji,k}\right)$ the two contributions can sum up to

$$3\cdot 2\sum_{\{3\}}\left[\left(u_i^2 u_j h_{ij,k}\right)+\left(u_i^2 u_k h_{ik,j}\right)+\left(u_j^2 u_k h_{jk,i}\right)\right] \qquad \text{(A24)}$$

and, because of the equality $2\sum_{\{3\}}\left[(ij,k)+(ki,j)+(jk,i)\right]=\sum_{\{1\}}\sum_{\{1\}\backslash i}\sum_{\{1\}\backslash i,j}(ij,k)$, eqn. A24 becomes

$$3\sum_{\{1\}}\sum_{\{1\}\backslash i}\sum_{\{1\}\backslash i,j}\left(u_i^2 u_j h_{ij,k}\right). \qquad \text{(A25)}$$

Similarly, the term $3\cdot 2\sum_{\langle 2\rangle} u_i u_i \sum_{\langle 1\rangle} u_j$ together with the analogous terms $(ik)$ e $(jk)$ lead to

$$3\sum_{\{1\}}\sum_{\{1\}\setminus i}\sum_{\{1\}\setminus i,j}\left(u_i u_i u_j h_{iij,k}\right). \tag{A26}$$

The forth line yields

$$6\sum_{\{3\}}\left(u_i u_j u_k h_{ijk}\right)=\sum_{\{1\}}\sum_{\{1\}\setminus i}\sum_{\{1\}\setminus i,j}\left(u_i u_j u_k h_{ijk}\right) \tag{A27}$$

Summing up eqns. A23 and A25-A27 one gets

$$\begin{aligned}\left\langle U^{(3)^3}\right\rangle_{\{3\}} &= \sum_{\{1\}}\sum_{\{2\}\setminus i}\left[\left(u_i^3 h_{i,jk}\right)+3\left(u_i^2 u_i h_{ii,jk}\right)+\left(u_i u_i u_i h_{iii,jk}\right)\right]+\\ &+\sum_{\{1\}}\sum_{\{1\}\setminus i}\sum_{\{1\}\setminus i,j}\left[3\left(u_i^2 u_j h_{ij,k}\right)+3\left(u_i u_i u_j h_{iij,k}\right)+\left(u_i u_j u_k h_{ijk}\right)\right],\end{aligned} \tag{A28}$$

and combining eqns.A18,A22 and A28 the average of $U^3$ is at last obtained

$$\begin{aligned}\left\langle U^3\right\rangle &= \sum_{\{1\}}\left[\left(u_i^3 h_i\right)+3\left(u_i^2 u_i h_{ii}\right)+\left(u_i u_i u_i h_{iii}\right)\right]+\\ &+\sum_{\{1\}}\sum_{\{1\}\setminus i}\left[\left(u_i^3 h_{i,j}\right)+3\left(u_i^2 u_i h_{ii,j}\right)+\left(u_i u_i u_i h_{iii,j}\right)+3\left(u_i^2 u_j h_{ij}\right)+3\left(u_i u_i u_j h_{iij}\right)\right]+\\ &+\sum_{\{1\}}\sum_{\{2\}\setminus i}\left[\left(u_i^3 h_{i,jk}\right)+3\left(u_i^2 u_i h_{ii,jk}\right)+\left(u_i u_i u_i h_{iii,jk}\right)\right]+\\ &+\sum_{\{1\}}\sum_{\{1\}\setminus i}\sum_{\{1\}\setminus i,j}\left[3\left(u_i^2 u_j h_{ij,k}\right)+3\left(u_i u_i u_j h_{iij,k}\right)+\left(u_i u_j u_k h_{ijk}\right)\right]=\\ &= \sum_{\{1\}}\left\{\left[\left(u_i^3 (h_i+\sum_{\{1\}\setminus i} h_{i,j}+\sum_{\{2\}\setminus i} h_{i,jk}+\ldots)\right)+\left(u_i^2 u_i (3h_{ii}+3\sum_{\{1\}\setminus i} h_{ii,j}+3\sum_{\{2\}\setminus i} h_{ii,jk}+\ldots)\right)+\right.\right.\\ &+\left.\left(u_i u_i u_i (h_{iii}+\sum_{\{1\}\setminus i} h_{iii,j}+\sum_{\{2\}\setminus i} h_{iii,jk}+\ldots)\right)+\right.\\ &\left.+3\sum_{\{1\}\setminus i}\left[\left(u_i^2 u_j (h_{ij}+\sum_{\{1\}\setminus i,j} h_{ij,k}+\ldots)\right)+\left(u_i u_i u_j (h_{iij}+\sum_{\{1\}\setminus i,j} h_{iij,k}+\ldots)\right)\right]+\sum_{\{1\}\setminus i}\sum_{\{1\}\setminus i,j}\left(u_i u_j u_k (h_{ijk}+\ldots)\right)\right\}=\\ &= \sum_{\{1\}}\left\{\left(u_i^3 f_i\right)+3\left(u_i^2 u_i f_{ii}\right)+\left(u_i u_i u_i f_{iii}\right)+3\sum_{\{1\}\setminus i}\left[\left(u_i^2 u_j f_{ij}\right)+\left(u_i u_i u_j f_{iij}\right)\right]+\sum_{\{1\}\setminus i}\sum_{\{1\}\setminus i,j}\left(u_i u_j u_k, f_{ijk}\right)\right\}.\end{aligned} \tag{A29}$$

In writing the last term of eqn.A29 the following $f$-functions have been defined for any tern of classes as

$$f_{iii} = h_{iii} + \sum_{\{1\}\setminus i} h_{iii,j} + \sum_{\{2\}\setminus i} h_{iii,jk} + \ldots \tag{A30}$$

$$f_{iij} = h_{iij} + \sum_{\{1\}\setminus i,j} h_{iij,k} + \ldots \tag{A31}$$

$$f_{ijk} = h_{ijk} + \ldots \tag{A32}$$

It is easy to verify that eqn.A29 can be rewritten in a more compact form as

$$\langle U^3 \rangle = \sum_{\{1\}} \left( u_i^3 f_i \right) + 3 \sum_{\{1\}} \sum_{\{1\}} \left( u_i^2 u_j f_{ij} \right) + \sum_{\{1\}} \sum_{\{1\}} \sum_{\{1\}} \left( u_i u_j u_k f_{ijk} \right). \tag{A33}$$

**Appendix B**

In case of a single class eqn.13 becomes $\left\langle \prod^{n_i} (1+u_i) \right\rangle = \left\langle 1 + \sum_{\langle 1 \rangle} u_i + \sum_{\langle 2 \rangle} u_i u_i + \sum_{\langle 3 \rangle} u_i u_i u_i + \ldots \right\rangle$ and by exploiting the results obtained in calculating the averages of the variable $U$ and its powers, it is possible to verify that the average operation on the functional gives

$$\left\langle \sum_{\langle 1 \rangle} u_i \right\rangle = \sum_{\{1\}} \left( u_i h_i \right) \tag{B1'}$$

$$\left\langle \sum_{\langle 2 \rangle} u_i u_i \right\rangle = \frac{1}{2!} \sum_{\{1\}} \left( u_i u_i h_{ii} \right) \tag{B2'}$$

$$\left\langle \sum_{\langle 3 \rangle} u_i u_i u_i \right\rangle\rangle \ \langle = \frac{1}{3!} \sum_{\{1\}} \left( u_i u_i u_i h_{iii} \right) \tag{B3'}$$

In case of two classes eqn.13 reduces to;

;

$$\left\langle \prod_{\langle 1\rangle}^{n_i}(1+u_i)\prod_{\langle 1\rangle}^{n_j}(1+u_j)\right\rangle = \left\langle 1+\sum_{\langle 1\rangle}u_i+\sum_{\langle 1\rangle}u_j+\sum_{\langle 2\rangle}u_iu_i+\sum_{\langle 2\rangle}u_ju_j+\sum_{\langle 1\rangle}\sum_{\langle 1\rangle}u_iu_j+\right.$$

$$\left.+\sum_{\langle 3\rangle}u_iu_iu_i+\sum_{\langle 3\rangle}u_ju_ju_j+\sum_{\langle 2\rangle}\sum_{\langle 1\rangle}u_iu_iu_j+\sum_{\langle 1\rangle}\sum_{\langle 2\rangle}u_iu_ju_j+\ldots\right\rangle$$

and averaging

$$\left\langle \sum_{\langle 1\rangle}u_i+\sum_{\langle 1\rangle}u_j\right\rangle = \sum_{\{1\}}\sum_{\{1\}\backslash i}\left(u_ih_{i,j}\right) \tag{B1''}$$

$$\left\langle \sum_{\langle 2\rangle}u_iu_i+\sum_{\langle 2\rangle}u_ju_j\right\rangle = \frac{1}{2!}\sum_{\{1\}}\sum_{\{1\}\backslash i}\left(u_iu_ih_{ii,j}\right) \tag{B2''}$$

$$\left\langle \sum_{\langle 3\rangle}u_iu_iu_i+\sum_{\langle 3\rangle}u_ju_ju_j\right\rangle = \frac{1}{3!}\sum_{\{1\}}\sum_{\{1\}\backslash i}\left(u_iu_iu_ih_{iii,j}\right) \tag{B3''}$$

$$\left\langle \sum_{\langle 1\rangle}\sum_{\langle 1\rangle}u_iu_j\right\rangle = \frac{1}{2}\sum_{\{1\}}\sum_{\{1\}\backslash i}\left(u_iu_jh_{ij}\right) = \sum_{\{2\}}\left(u_iu_jh_{ij}\right) \tag{B4''}$$

$$\left\langle \sum_{\langle 2\rangle}\sum_{\langle 1\rangle}u_iu_iu_j+\sum_{\langle 1\rangle}\sum_{\langle 2\rangle}u_iu_ju_j\right\rangle = \frac{1}{2!}\sum_{\{1\}}\sum_{\{1\}\backslash i}\left(u_iu_iu_jh_{iij}\right) \tag{B5''}$$

In case of three classes eqn.13 reduces to $\left\langle \prod_{\langle 1\rangle}^{n_i}(1+u_i)\prod_{\langle 1\rangle}^{n_j}(1+u_j)\prod_{\langle 1\rangle}^{n_k}(1+u_k)\right\rangle =$

$$= \left\langle 1+\sum_{\langle 1\rangle}u_i+\sum_{\langle 1\rangle}u_j+\sum_{\langle 1\rangle}u_k+\sum_{\langle 2\rangle}u_iu_i+\sum_{\langle 2\rangle}u_ju_j+\sum_{\langle 2\rangle}u_ku_k+\sum_{\langle 1\rangle}\sum_{\langle 1\rangle}u_iu_j+\sum_{\langle 1\rangle}\sum_{\langle 1\rangle}u_iu_k+\sum_{\langle 1\rangle}\sum_{\langle 1\rangle}u_ju_k+\right.$$

$$+\sum_{\langle 3\rangle}u_iu_iu_i+\sum_{\langle 3\rangle}u_ju_ju_j+\sum_{\langle 3\rangle}u_ku_ku_k+\sum_{\langle 1\rangle}\sum_{\langle 2\rangle}u_iu_iu_j+\sum_{\langle 1\rangle}\sum_{\langle 2\rangle}u_iu_iu_k+\sum_{\langle 1\rangle}\sum_{\langle 2\rangle}u_ju_ju_i+\sum_{\langle 1\rangle}\sum_{\langle 2\rangle}u_ju_ju_k+$$

$$\left.+\sum_{\langle 1\rangle}\sum_{\langle 2\rangle}u_ku_ku_j+\sum_{\langle 1\rangle}\sum_{\langle 2\rangle}u_ku_ku_i+\sum_{\langle 1\rangle}\sum_{\langle 1\rangle}\sum_{\langle 1\rangle}u_iu_ju_k+\ldots\right\rangle$$

and averaging

$$\left\langle \sum_{\langle 1\rangle}u_i+\sum_{\langle 1\rangle}u_j+\sum_{\langle 1\rangle}u_k\right\rangle = \sum_{\{1\}}\sum_{\{2\}\backslash i}\left(u_ih_{i,jk}\right) \tag{B1'''}$$

$$\left\langle \sum_{\langle 2\rangle}u_iu_i+\sum_{\langle 2\rangle}u_ju_j+\sum_{\langle 2\rangle}u_ku_k\right\rangle = \frac{1}{2}\sum_{\{1\}}\sum_{\{2\}\backslash i}\left(u_iu_ih_{ii,jk}\right) \tag{B2'''}$$

$$\left\langle \sum_{\langle 3\rangle} u_i u_i u_i + \sum_{\langle 3\rangle} u_j u_j u_j + \sum_{\langle 3\rangle} u_k u_k u_k \right\rangle = \frac{1}{3!}\sum_{\{1\}}\sum_{\{2\}\backslash i}\left(u_i u_i u_i h_{iii,jk}\right) \qquad \text{(B3''')}$$

$$\left\langle \sum_{\langle 1\rangle}\sum_{\langle 1\rangle} u_i u_j + \sum_{\langle 1\rangle}\sum_{\langle 1\rangle} u_i u_k + \sum_{\langle 1\rangle}\sum_{\langle 1\rangle} u_j u_k \right\rangle = \frac{1}{2}\sum_{\{1\}}\sum_{\{1\}\backslash i}\sum_{\{1\}\backslash i,j}\left(u_i u_j h_{ij,k}\right) = \sum_{\{3\}}\left(u_i u_j h_{ij,k}\right) \qquad \text{(B4''')}$$

$$\left\langle \sum_{\langle 1\rangle}\sum_{\langle 2\rangle} u_i u_i u_j + \sum_{\langle 1\rangle}\sum_{\langle 2\rangle} u_i u_i u_k + \ldots + \sum_{\langle 1\rangle}\sum_{\langle 2\rangle} u_k u_k u_i \right\rangle = \frac{1}{2}\sum_{\{1\}}\sum_{\{1\}\backslash i}\sum_{\{1\}\backslash i,j}\left(u_i u_i u_j h_{iij,k}\right) \qquad \text{(B5''')}$$

$$\left\langle \sum_{\langle 1\rangle}\sum_{\langle 1\rangle}\sum_{\langle 1\rangle} u_i u_j u_k \right\rangle = \frac{1}{6}\sum_{\{1\}}\sum_{\{1\}\backslash i}\sum_{\{1\}\backslash i,j}\left(u_i u_j u_k h_{ijk}\right) \qquad \text{(B6''')}$$

It is recognised that, once summed, eqns.B1 lead to $\sum_{\{1\}}\left(u_i f_i\right)$, eqns.B2 lead to $\frac{1}{2!}\sum_{\{1\}}\left(u_i u_i f_{ii}\right)$, eqns.B3 lead to $\frac{1}{3!}\sum_{\{1\}}\left(u_i u_i u_i f_{iii}\right)$, eqns.B4 lead to $\frac{1}{2}\sum_{\{1\}}\sum_{\{1\}\backslash i}\left(u_i u_j f_{ij}\right) = \sum_{\{2\}}\left(u_i u_j f_{ij}\right)$, eqns.B5 lead to $\frac{1}{2!}\sum_{\{1\}}\sum_{\{1\}\backslash i}\left(u_i u_i u_j f_{iij}\right)$ and eqns.B6 lead to $\frac{1}{3!}\sum_{\{1\}}\sum_{\{1\}\backslash i}\sum_{\{1\}\backslash i,j}\left(u_i u_j u_k f_{ijk}\right) = \sum_{\{3\}}\left(u_i u_j u_k f_{ijk}\right)$. In other words, the average of the functional eqn.13 becomes

$$L[\{u\}] = 1 + \sum_{\{1\}}\left(u_i f_i\right) + \frac{1}{2}\sum_{\{1\}}\left(u_i u_i f_{ii}\right) + \frac{1}{3!}\sum_{\{1\}}\left(u_i u_i u_i f_{iii}\right) + \sum_{\{2\}}\left(u_i u_j f_{ij}\right) + \frac{1}{2!}\sum_{\{1\}}\sum_{\{1\}\backslash i}\left(u_i u_i u_j f_{iij}\right) + \sum_{\{3\}}\left(u_i u_j u_k f_{ijk}\right)$$

........ (B7)

It is possible to rewrite eqn.(B7) in a more familiar and manageable form as reported in eqn. 14

**Appendix C**

The equivalence between eqns. 33 and 35, has been exploited in ref.1 for the case m=2. Here we will show it for the case m=3. For m >3 can be followed the same path of computation. Let us single out the contribution s=3 from eqn.33. It is

$$T_3 = \sum_{m=1}^{3}\sum_{\{m\}}\sum_{\Pi_m^3}\frac{-I_3^{\{m\}}}{\boldsymbol{n}_1^3!..\boldsymbol{n}_m^3!} = -\left(\sum_{\{1\}}\frac{I_3^{(1)}}{3!} + \sum_{\{2\}}\left[\frac{I_3^{(2,1)}}{2!1!} + \frac{I_3^{(1,2)}}{2!1!}\right] + \sum_{\{3\}}\frac{I_3^{(3)}}{1!1!1!}\right) \qquad \text{(C1)}$$

that, for the sake of computation convenience, can be rewritten as

$$T_3 = -\left(\sum_i \frac{I_3^{(i,i,i)}}{3!} + \sum_{i>j}\left[\frac{I_3^{(i,i,j)}}{2!} + \frac{I_3^{(j,j,i)}}{2!}\right] + \sum_{i>j>k} I_3^{(i,j,k)}\right) = -\left(\sum_i \frac{I_3^{(i,i,i)}}{3!} + \sum_{i\neq j}\left[\frac{I_3^{(i,i,j)}}{2!}\right] + \frac{1}{3!}\sum_{i\neq j\neq k} I_3^{(i,j,k)}\right)$$

(C2)

Equation 35 yields, for s=3,

$$T'_3 = \frac{1}{3!}\sum_{i,j,k} I_3^{(i,j,k)} = \frac{1}{3!}\left[\sum_i I_3^{(i,i,i)} + 3\sum_{i\neq j} I_3^{(i,i,j)} + \sum_{i\neq j\neq k} I_3^{(i,j,k)}\right] \quad \text{(C3)}$$

apparently $T_3 = T'_3$ .